\documentstyle[prd,aps,epsf]{revtex}

\begin{document}

\twocolumn[\hsize\textwidth\columnwidth\hsize\csname
@twocolumnfalse\endcsname

\title{Uniqueness and non-uniqueness of static black holes in higher dimensions}

\author{${}^{(a)}$Gary W. Gibbons, ${}^{(b)}$Daisuke Ida and
${}^{(b)}$Tetsuya Shiromizu}

\address{${}^{(a)}$DAMTP, Centre for Mathematical Sciences, The University
of Cambridge,\\
Wilberforce Road, Cambridge CB3 0WA, United Kingdom \\
${}^{(b)}$Research Center for the Early Universe (RESCEU),
The University of Tokyo, Tokyo 113-0033, Japan}

\maketitle

\begin{abstract}
We prove a  uniqueness theorem for
 asymptotically flat static charged dilaton black
hole
solutions in higher dimensional space-times. We also construct
infinitely many non-asymptotically flat regular static black holes
on the same space-time manifold with the same spherical topology.
An application to the uniqueness of a certain class of  flat $p$-branes is also
given.\end{abstract}
\vskip2pc]

\vskip1cm

With the development of string theory, black holes in higher dimensional
space-times have come to play a fundamental role in physics \cite{BH}.
Furthermore,
the possibility of black hole production
in high energy experiments has  recently been suggested  in the context of
the
so-called TeV gravity \cite{LHC}.
A TeV-size black hole in TeV gravity is small enough to be well approximated
by
an asymptotically flat black hole in higher dimensions.
To predict  phenomenological
results, we need reliable  knowledge about higher dimensional black
holes.
However, some essential  features of black hole
theory have not so far been fully explored in many
recent  discussions.
Among these, the equilibrium problem for  black holes is
one of most important issues.
The final equilibrium state of the black hole is known to
drastically simplify  in the case of four space-time dimensions
because of the uniqueness properties  of
static or stationary black hole solutions.
The uniqueness theorem for  the vacuum black hole has been well established
in four-dimensional space-times \cite{4d1,4d2,4d3} (See also
Ref.~\cite{Review} for
comprehensive review).
Although this no hair property is  fundamental to the nature of black holes,
it is at the same time a quite non-trivial result derived from the Einstein
equations.
Remarkably, five-dimensional stationary vacuum black holes are  not unique;
there is a Myers-Perry solution \cite{BH}, which is a
generalization of the Kerr solution
to arbitrary dimensions, while Emparan and Reall \cite{Reall} have recently
found  five-dimensional rotating black ring solutions with the same
angular momenta and mass but now the  event horizon
homeomorphic to $S^2\times S^1$.
In the static case, 
such a counter-example has not yet been presented.
The only known asymptotically flat static vacuum black hole
is the $n$-dimensional hyperspherically symmetric
Schwarzschild-Tangherlini solution \cite{Tangherlini}.
In this paper we shall show that there are no others.
 However it is interesting to note that, as we shall expand upon below,
if one drops the condition of asymptotic flatness but still
insists that the space-time has the same topology as that of the
Schwarschild-Tanghelerlini solution, then the uniqueness property
fails badly. There exist discrete infinities  of solutions.

In this paper we shall show that static  charged dilatonic
non-extreme black holes with a certain dilaton coupling constant are also
unique. 
(See also \cite{Hwang} for an   earlier paper  on the vacuum case
in the context of Riemannian geometry).  

Our work was  motivated by models in 
string theory where  gauge fields often play an essential role. 
Another important motivation is to embark on the programme
of proving the uniqueness of static black $p$-brane solutions.
These are $(n+p)$-dimensional  spacetimes invariant under the action of a 
$p$-dimensional Abelian translation group. Reduction to $n$ spacetime
dimensions produces a black hole solution of gravity coupled
to one or more scalars and an electric $2$-form  or dual
magnetic  $(n-2)$-form field strength. We  shall return to this
aspect of our work in a later publication.

The main tools we use are    method developed by Masood-ul-Alam\cite{4emd} 
to deal with charged dilaton black holes in 3+1 spacetime dimensions
and a new method of our own (introduced independently by
\cite{Hwang})
 based on
the use of totally umbillic hypersurfaces.
The latter is an essential ingredient becaues existing 3+1 methods do
apply.
It is also essential to assume strict asymptotic flatness
because using some Einstein  metrics on $S^{n-2}$,
$n=7,\dots, 11$ due to Bohm we can construct infinitely many
non-asymptotically flat solutions on manifolds wuth the same topology.
In addition, just as in 3+1 dimensions,
we also have to assume that the surface
gravity is non-zero (non-extreme), otherwise one has multi-black holes 
solutions.

The first half of our work  parallels that of   Bunting and
Masood-ul-Alam
\cite{4d2}, so that we shall only  briefly describe it  here.

We start with the Lagrangian in $n$-dimensional space-time
\begin{equation}
L={}^nR-2(\partial\phi)^2-e^{-\alpha\phi}F^2,
\end{equation}
where $F$ is the Maxwell field and the dilaton coupling constant is set to
$\alpha=[8(n-3)/(n-2)]^{1/2}$\footnote{Note that our proof presented here is only 
true for this choice for $\alpha$. If not, for example, Eqs (\ref{bad1}), 
(\ref{divergence}) and 
(\ref{bad3}) are not correct and then proof is not correct.} 
In general, the metric of an  $n$-dimensional
static space-time has the form
%
\begin{eqnarray}
ds^2=-V^2dt^2+g_{ij}dx^idx^j,
\end{eqnarray}
%
where $V$ and $g_{ij}$ are independent of $t$
and they are regarded as quantities on the $t={\rm const.}$
hypersurface $\Sigma$.
The event horizon $H$ is a Killing horizon
located at the level set $V=0$,
which is assumed to be non-degenerate. In fact non-degeneracy  follows from
Smarr's formula relating the mass, surface gravity and area of the horizon.
Then the static field equations become
%
\begin{eqnarray}
\nabla^2V&=&{\alpha^2e^{-\alpha\phi}\over16V}(\nabla\psi)^2, \label{einstein1}\\
\nabla^2\phi&=&{\alpha e^{-\alpha\phi}\over8V^2}(\nabla\psi)^2
-{\nabla V\cdot\nabla\phi\over V}, \label{dilaton} \\
\nabla^2\psi&=&\nabla\psi\cdot\left(\alpha\nabla\phi+{\nabla V\over V}\right),
\label{Maxwell}
\end{eqnarray}
and
\begin{eqnarray}
R_{ij}&=&{\nabla_i\nabla_j V\over V}+2\nabla_i\phi\nabla_j\phi\nonumber\\
&&{}-{e^{-\alpha\phi}\over 2V^2}\left[
\nabla_i\psi\nabla_j\psi-{(\nabla\psi)^2\over n-2}g_{ij}
\right], \label{einstein2}
\end{eqnarray}
%
where $\nabla$ and $R_{ij}$ denote  covariant derivative
and the Ricci tensor defined on
$(\Sigma, g_{ij})$, respectively, and $\psi$ is the electrostatic potential
such that $F=d\psi\wedge dt$.

In asymptotically
flat space-times, one can find an appropriate coordinate system in which
the metric, dilaton and electrostatic potential  have  asymptotic expansions of the form
\begin{eqnarray}
V &=& 1-\frac{\mu}{r^{n-3}}+O(1/r^{n-2}),\\
g_{ij}&=&\left(
1+\frac{2}{n-3}\frac{\mu}{r^{n-3}}\right)\delta_{ij}+O(1/r^{n-2}),\\
e^{\alpha\phi/2}&=&1-{Q^2\over\mu r^{n-3}}+O(1/r^{n-2}),\\
\psi&=&{2^{5/2}Q\over\alpha r^{n-3}}+O(1/r^{n-2}),
\end{eqnarray}
respectively,
where $\mu$, $Q={\rm const.}$ represent the ADM mass 
and the electric charge (up to  constant factors), respectively,
and $r:=\sqrt{\sum_i(x^i)^2}$. We assume the non-extremal condition
 $\mu>|Q|$.

Consider the following two conformal transformations
\begin{eqnarray}
\hat g_{ij}^{\pm}=(e^{-\alpha\phi/2}\lambda_1^\pm\lambda_2^\pm)^{4/(n-3)} g_{ij},
\end{eqnarray}
where
\begin{eqnarray}
\lambda_1^\pm&:=&{1\pm Ve^{\alpha\phi/2}\over 2}+2^{-5/2}\alpha\psi,\\
\lambda_2^\pm&:=&{1\pm Ve^{\alpha\phi/2}\over 2}-2^{-5/2}\alpha\psi.\\
\end{eqnarray}
Then we have two manifolds ($ \Sigma^{\pm}, g_{ij}^{\pm}$).
On $\Sigma^{+}$, the asymptotic behavior of the metric
becomes
\begin{eqnarray}
\hat g_{ij}^{+} =\delta_{ij}+ O\left(1/r^{n-2} \right).\label{asymptotic}
\end{eqnarray}
On  $ \Sigma^{-}$, we have
\begin{eqnarray}
\hat g_{ij}^{-} dx^i dx^j & = &
\left({\mu^2-Q^2\over2\mu}\right)^{4/(n-3)}
\left(d\varrho^2+\varrho^2d\Omega^2_{n-2}\right)\nonumber\\
&&{}+O(\varrho^5).
\end{eqnarray}
where $d\Omega^2_{n-2}$ denotes the round sphere metric and $\varrho:=1/r$ has
been defined.
Pasting $( \Sigma^{\pm}, g_{ij}^{\pm})$ across the
level set $V=0$ and adding a point $\{p\}$ at $\varrho=0$,
we can construct a complete regular surface
$\hat\Sigma= \Sigma^{+} \cup \Sigma^{-}\cup \{p\}$.
The Ricci scalar (sometimes called the  scalar curvature)
 $\hat R$ on $\Sigma^\pm$ becomes
\begin{eqnarray}
\hat R
&=&{4\over\alpha^2}\Biggl|
\hat\nabla\ln{\lambda_1^\pm\over\lambda_2^\pm}
\mp{\alpha e^{-\alpha\phi/2}\over2^{3/2}V} \hat\nabla\psi\Biggr|^2
+{16\over\alpha^2}{\hat\nabla^2\tau_\pm\over\tau_\pm}, \label{bad1}
\label{ricci}
\end{eqnarray}
where $\tau_\pm:=e^{\alpha\phi/2}(\lambda_1^\pm\lambda_2^\pm)^{-1/2}$
has been defined. Note that the last term in (\ref{ricci}) has no
definite sign. Nevertheless we can  still generalize the standard
positive energy theorem due to Witten to cover this new situation.

Following \cite{4emd}  we consider the Witten spinor $\Psi$ 
(an asymptotically constant spinor satisfying $\gamma^iD_i\Psi=0$) 
on ($\hat\Sigma^\pm,g^\pm_{ij}$). Then we have a divergence identity
\begin{eqnarray}
&&\hat\nabla\cdot\left[
\hat\nabla|\Psi|^2
-{8\over\alpha^2}|\Psi|^2 \hat\nabla\ln{\tau_\pm\over\sigma_\pm}
\right]\nonumber\\
&=&
\left[
{2\over\alpha^2}\Biggl|
\hat\nabla\ln{\lambda_1^\pm\over\lambda_2^\pm}
\mp {\alpha e^{-\alpha\phi/2}\over 2^{3/2} V} \hat\nabla\psi
\Biggr|^2\right.\nonumber\\
&&{}\left.+{(n-2)(n-4)\over 2(n-3)^2}\Biggl|
\hat\nabla\ln{\tau_\pm\over\sigma_\pm}
\Biggr|^2\right]|\Psi|^2\nonumber\\
&&+2\Biggl|
D\Psi-{4\over\alpha^2}\left(\hat\nabla\ln{\tau_\pm\over\sigma_\pm}\right)\Psi
\Biggl|^2,
\label{divergence}
\end{eqnarray}
where $\sigma_\pm:=(1\pm Ve^{-\alpha\phi/2})^{-1}$ has been defined.
We have used
\begin{eqnarray}
\hat\nabla^2|\Psi|^2&=&{\hat R\over 2}|\Psi|^2+2|D\Psi|^2,\\
{\hat\nabla^2\sigma_\pm\over\sigma_\pm}&=&2{\hat\nabla\sigma_\pm\over\sigma_\pm}\cdot
\left({\hat\nabla\sigma_\pm\over\sigma_\pm}-{\hat\nabla\tau_\pm\over\tau_\pm}\right)
\label{bad3}
\end{eqnarray}
Integrating Eq.~(\ref{divergence}) over $\hat\Sigma$, 
 positive mass theorem
can be proven.
In other words, the l.h.s.  represents the total gravitational mass on $\hat\Sigma$ 
and it is positive definite from the expression in the r.h.s..
However, Eq.~(\ref{asymptotic}) implies that the total  gravitational mass is zero on
$\hat\Sigma$. This means that the each terms in the r.h.s of Eq.~(\ref{divergence}) vanish:
\begin{eqnarray}
\hat\nabla_i\ln{\lambda_1^\pm\over\lambda_2^\pm}
&=&\pm {\alpha e^{-\alpha\phi/2}\over 2^{3/2} V} \hat\nabla_i\psi,
\label{rhs1}\\
\hat\nabla_i\ln{\tau_\pm\over\sigma_\pm}&=&0,
\label{rhs2}\\
D_i\Psi&=&0.\label{rhs3}
\end{eqnarray}
Equations~(\ref{rhs1}), (\ref{rhs2}) imply that the level sets of
$V$, $\phi$ and $\psi$ conincide.
Furthermore, the spinor equation~(\ref{rhs3}) implies that $\hat\Sigma$ is the flat space: $\hat g_{ij}=\delta_{ij}$
Then, Eq.~(\ref{ricci}) implies that $\tau_\pm$ is a harmonic function on the flat space:
$\nabla_0^2\tau_\pm=0$, where $\nabla_0$ denotes the flat connection.

We shall now demonstrate that the conformally transformed event horizon 
$\hat H$ is a geometric sphere in $\hat\Sigma$.
We choose $V$ as a local coordinate in a neighbour hood $U \subset \Sigma$.
Let $\{x^A\}$ be coordinates on level sets of $V$
such that their trajectries are orthogonal to each level set.
Then, the metric on $\Sigma$ can be  written in the form
\begin{equation}
g=\rho^2dV^2+h_{AB}dx^Adx^B,
\end{equation}
where $\rho^2:=(\nabla V)^2$. 
Since $\Sigma$ is conformally flat, the Riemann invariant has
a simple expression in this coordinate system:
\begin{eqnarray}
{}^nR_{IJKL}{}^nR^{IJKL}&=&R_{ijkl}R^{ijkl}+4R_{0i0j}R^{0i0j}\nonumber\\
&  = & \frac{4(n-2)}{(n-3)V^2\rho^2}\left[k_{AB}k^{AB}+k^2\right.\nonumber\\
&&\left.{}+2 {\cal D}_A \rho
{\cal D}^A \rho\right],
\label{kretchmann}
\end{eqnarray}
where ${\cal D}_A$ denotes the covariant derivative on each level set of
$V$, and $k_{AB}$ is the second fundamental form of the level set.

The requirement that the event horizon $H$ is a regular surface
leads to the condition
\begin{eqnarray}
&&k_{AB}|_H=0,\\
&&{\cal D}_A\rho|_H=0.
\end{eqnarray}
In particular, $H$ is a totally geodesic surface in $\Sigma$.

Let us consider how the event horizon
is embedded into the base space $(\hat\Sigma,\delta_{ij})$.
In terms of the smooth function $\tau:=\tau_\pm$, we can 
adopt the following local expression for the flat space
\begin{eqnarray}
\delta_{ij}dx^idx^j=\hat \rho^2 d\tau^2+\hat h_{AB}dx^A dx^B.
\end{eqnarray}
The event horizon is located at some $\tau=\tau_H={\rm const.}$ surface $\hat H$.
Then, the extrinsic curvature $\hat k_{AB}$ of the level set $\tau={\rm
const.}$
can be expressed as
\begin{eqnarray}
\hat k_{AB}&=&e^{\alpha\phi/(n-3)}\tau^{-4/(n-3)}k_{AB}\nonumber\\
&&{}+\frac{2}{n-3}
\frac{(e^{-\alpha\phi/4}\tau)^{2(n-1)/(n-3)}}{\rho} \hat h_{AB}.
\end{eqnarray}
Thus we obtain
\begin{equation}
\hat k_{AB}=\frac{2}{n-3}
\frac{(e^{-\alpha\phi/4}\tau)^{2(n-1)/(n-3)}}{\rho}\Biggr|_H \hat h_{AB}.
\end{equation}
on $\hat H$. In other words, the embedding of $\hat H$ into the
Euclidean $(n-1)$-space
is totally umbilical.
It is known that such a embedding must be hyperspherical \cite{sphere},
namely
each connected component of $\hat H$ is a geometric sphere with a certain
radius.
The embedding of a hypersphere into the Euclidean space is known to be rigid
\cite{rigid},
which means that we can always locate one connected component of
 $\tilde H$ at the $r=r_0$ surface of $\tilde \Sigma$
without loss of generality. If there is only a single horizon, we have
a boundary value problem for  the
 Laplace equation $\nabla_0^2\tau=0$ on the base space
$\Omega:=E^{n-1}\setminus B^{n-1}$ with the Dirichlet boundary conditions.
Such a solution must be spherically symmetric, so that it is given by
the solutions found in  \cite{GM}.

One may remove the assumption of the single horizon as follows.
Consider the evolution of the level surface in
Euclidean space. From the Gauss equation in Euclidean space one
obtains  the evolution equation for the shear $\hat \sigma_{AB}
:=\hat k_{AB}-\hat k \hat h_{AB}/(n-2)$:
\begin{eqnarray}
\mbox \pounds_{\hat n}  \hat \sigma_{AB}
& = & \hat \sigma_A{}^C \hat \sigma_{CB}
+\frac{1}{n-2}\hat h_{AB} \hat \sigma_{CD} \hat \sigma^{CD} \nonumber
\\
& & -\frac{1}{\hat \rho}\biggl( {\hat {\cal D}}_A {\hat {\cal D}}_B
-\frac{1}{n-2}\hat h_{AB} {\hat {\cal D}}^2  \biggr)\hat \rho,
\end{eqnarray}
where $\hat n$ denotes the unit normal to the level set of $\tau$.
Using $\nabla_0^2 \tau=0$, we obtain
\begin{eqnarray}
\mbox \pounds_{\hat n} {\hat {\cal D}}_A {\rm ln}\hat \rho&=&
\hat k {\hat {\cal D}}_A {\rm ln}\hat \rho +{\hat {\cal D}}_A \hat
k,\\
\mbox \pounds_{\hat n} \hat k &=& -
||\hat\sigma||^2-\frac{1}{n-2}k^2-\frac{1}{\hat \rho}
{\hat {\cal D}}^2 \hat \rho,\\
\mbox \pounds_{\hat n} {\cal D}_A \hat k &=&{\hat {\cal D}}_A \mbox
\pounds_{\hat n}
\hat k + ({\hat {\cal D}}_A {\rm ln}\hat \rho)(  \mbox \pounds_{\hat
n}
\hat k).
\end{eqnarray}
From the above equations, it can be seen that
\begin{eqnarray}
\hat \sigma_{AB}=0, ~~~
{\hat {\cal D}}_A \hat \rho=0,~~~
{\hat {\cal D}}_A\hat k =0,
\end{eqnarray}
that is, each level surface of $v$ is totally
umbilic and hence spherically symmetric, which implies
that the metric is isometric to those found in \cite{GM}.

This is of course local result since we consider only the region
containing no saddle points of the harmonic function $\tau$.
To obtain the global result, we need a further assumption such as
analyticity.
However, the assumption that there is no saddle point may be  justified as
follows.
At a saddle point $\rho=0$, the level surface of $\tau$ is multi-sheeted,
that is
the embedding of the level surfaces is singular there.
One can find at least one level surface such that $k_{AB}\ne 0$ near the
saddle point.
Then, Eq.~(\ref{kretchmann}) implies that the saddle point is singular.
If the horizon is not connected,
this naked singularity must  exist to
 compensate for the gravitational attraction
between black holes.

One motivation for the present work
was to embark on a programme of proving the uniqueness of static
$p$-brane solutions. In general these take the form
\begin{equation}
ds^2 = e^{2\gamma \Phi} ( d{\bf y}_p ^2 ) + e^ {2 \delta \Phi} g_{\mu
\nu} dx^\mu dx ^\nu. \label{metric5}
\end{equation}
Dimensional reduction on ${\bf E} ^p$ will take the Einstein-Hilbert
action to the Einstein-Hilbert action if $(n-2) \delta +p \gamma 
=0$. 
If the higher dimensional metric is coupled to an $n-2$ form $F_{n-2}$
with no scalars, we obtain the Lagrangian\cite{GHT} 
\begin{equation}
R -2 (\partial \phi)^2 - { 2 \over (n-2)! } e^{ -\alpha \phi} F^2 _{n-2}, 
\end{equation}
with 
\begin{equation}
\gamma^2 = { 2 (n-2) \over p(n+p-2) }
\end{equation}
and 
\begin{equation}
\alpha^2 = { 8 p(n-3)^2 \over (n-2) (n+p-2) }. 
\end{equation}
We can  now apply the results of this paper for 
$(n,p)=(6,2)~{\rm or}~(5,3)$. 
For magnetic branes  we need  to make a
duality transformation. We plan to give more details in a future publication.

We have shown that if space-time is assumed to asymptotically
flat then the only regular static ellectric dilaton 
black holes are hyper-spherically
symmetric and hence  given
the solutions in \cite{GM}. However other,
non-asymptotically flat, solutions may be obtained by replacing
the metric of the round $(n-2)$-sphere by any other Einstein manifold
whose Ricci-curvature has the same magnitude as that of a unit round
$(n-2)$-sphere. In particular we can replace the round metric on
$S^5,S^6,\dots, S^9$ by the infinite sequences of Einstein metrics
found recently by Bohm \cite{Bohm}. These have the form
\begin{eqnarray}
d\theta^2 + a^2(\theta ) g_p + b^2(\theta) g_q,
\end{eqnarray}
where $g_m$ is  the unit round
metric on the $m$-sphere, $p+q=n-3$ and neither $p$ nor $q$ is one.
In general these metrics are inhomogeneous with
isometry groups $SO(p+1)\times SO(q+1)$.
The round metric on $S^{n-2}$ is given by $a=\sin \theta$ and $b=\cos
\theta$. In addition, Bohm demonstrates the existence of infinite
sequences
of smooth Einstein  metrics which converge to  singular metrics of
finite
volume. By a theorem of Bishop \cite{Bishop}, the volume
of these metrics is always less than that of the round metric.
In the vacuum case,it follows that for fixed temperature, the associated static black
holes  always have smaller Bekenstein-Hawking entropy than the
spherical black hole.
For this reason we believe  that these metrics are all unstable.
Preliminary results of GWG obtained with  Sean Hartnoll 
show that the stability
depends upon the spectrum of the Lichnerowicz operator for the Bohm metrics.
This is not yet known. Presumably, if they are unstable, 
then  they will decay to  the spherical black-hole solution.
An interesting question, which is beyond the scope of the methods of
this paper because of the absence of a suitable 
positive energy theorem, is whether the Bohm black holes are unique
among metrics with the same asymptotics, i.e. those constructed from
the same Bohm
metric. 

One might question what physical relevance  the Bohm metrics, not being
asymptotically flat, 
might have. One possible application is that they
may be used in D3-branes solutions
and applied to the AdS/CFT correspondence.
The volume may then be related to
the central charge of a conformal field theory \cite{Gubser}.
In this way we obtain
a further connection between  the  entropy of horizons, geometry  and
the central charges of
quantum field theories where now Bishop's theorem provides a universal
bound for the central charge.

Another possibility is that in modeling scattering and in connection
with TeV black hole physics in the context of small extra dimensions,
the appropriate boundary condition may be more general
than the standard one of asymptotic flatness.

{\em Acknowledgments.}
We would like to thank H. Kodama and H. S. Reall
for useful discussion and comments.
TS's work is partially supported by Yamada Science Foundation.
GWG  would like to thank M. Anderson for bringing \cite{Hwang}
to his attention.


\begin{thebibliography}{22}

\bibitem{BH}
R.C. Myers and M.J. Perry, Ann. Phys. {\bf 172}, 304 (1986).


\bibitem{LHC}
P.C. Argyres, S. Dimopoulos and J. March-Russell, Phys. Lett.
{\bf B441}, 96 (1998);
%
R. Emparan, G.T. Horowitz and R.C. Myers,
Phys. Rev. Lett. {\bf 85}, 499 (2000);
%
S. Dimupoulos and G. Landsberg, Phys. Rev. Lett. {\bf 87}, 161602 (2001);
%
S.B. Giddings and S. Thomas, hep-th/0106219.

\bibitem{4d1}
W. Israel, Phys. Rev. {\bf 164}, 1776 (1967).

\bibitem{4d2}
G.L. Bunting and A.K.M. Masood-ul-Alam, Gen. Rel. Grav. {\bf 19}, 147
(1987).

\bibitem{4d3}
B. Carter, Phys. Rev. Lett. {\bf 26}, 331 (1971);
S.W. Hawking, Commun. Math. Phys. {\bf 25}, 152 (1972);
D.C. Robinson, Phys. Rev. Lett. {\bf 34}, 905 (1975).

\bibitem{Review}
B. Carter, 
and Black Hole Configurations.''
in {\it Gravitation in Astrophysics},
edited by B.~Carter and J.B. Hartle (Plenum, New York, 1987);
M.~Heusler, {\it Black Hole Uniqueness Theorems},
(Cambridge Univ. Press, London, 1996).

\bibitem{Reall}
R. Emparan and H.S. Reall, Phys. Rev. Lett. {\bf 88}, 101101 (2002).

\bibitem{Tangherlini}
F.R. Tangherlini, Nuovo Cimento {\bf 27}, 636 (1963).

\bibitem{sphere}
S. Kobayashi and K. Nomizu, {\it Foundations of Differential Geometry},
Vol.~II
(Interscience Publishers, New York, 1969),
Sec.~VII, Theorem~5.1.

\bibitem{rigid}
{\em ibid.} Theorem~6.4.
\bibitem{PET}
E. Witten, Commun. Math. Phys. {\bf 80}, 381 (1981).


\bibitem{Bohm} C. Bohm, Invent. Math {\bf 134}, 145-176 (1998).

\bibitem{Bishop} R. L. Bishop, Notices Amer Math Soc {\bf 10} 363
(1963).

\bibitem{Gubser} S. Gubser, Phys. Rev. D {\bf 59}, 025006 (1999).

\bibitem{Hwang}
S. Hwang, Geometriae Dedicata {\bf 71}, 5 (1998).



\bibitem{4emd}
A.K.M. Masood-ul-Alam, Class. Quantum Grav. {\bf 10}, 2649 (1993).

\bibitem{GM} G.W. Gibbons, Nucl Phys {\bf B207} 337 (1982),G. W. Gibbons and K. Maeda, Nucl Phys {\bf B298} 741 (1988)
\bibitem{GHT} G. W. Gibbons, G. Horowitz and P. K. Townsend, Class Quant
Grav {\bf 12} 297 (1995)
\end{thebibliography}
\end{document}